\documentclass[fleqn,twoside]{article}
\usepackage{espcrc2}
\usepackage{graphicx}
\usepackage{emp}
\newcommand{\empcopytoTeX}[1]{#1\empaddtoTeX{#1}}
\empcopytoTeX{\usepackage[latin1]{inputenc}}
\empcopytoTeX{\usepackage{amsmath,amssymb}}
\empcopytoTeX{}
\empaddtoprelude{input graph}
\empaddtoprelude{input sarith}
\empaddtoprelude{input rboxes}

\allowdisplaybreaks

\begin{document}
\title{Automated Calculation and Simulation Systems}
\author{Thorsten Ohl\address{%
  Lehrstuhl f\"ur theoretische Physik II,
  Universit\"at W\"urzburg,
  Am Hubland,
  97074 W\"urzburg, Germany
  \texttt{<ohl@physik.uni-wuerzburg.de>}}}
\begin{empfile}
\begin{abstract}
  I briefly summarize the parallel sessions on Automated Calculation
  and Simulation Systems for high energy particle physics
  phenomenology at ACAT\,2002 (Moscow State University, June 2002) and
  present a short overview over the current status of the field and
  try to identify the important trends.
\end{abstract}
\renewcommand{\rightmark}{WUE-ITP-2002-031}
\renewcommand{\leftmark}{WUE-ITP-2002-031}
\maketitle
\section{Introduction}
Future particle colliders---planned or already under
construction---will explore a new frontier in energy and precision.
Final states with more tagged particles and better defined jets will
become available for physics analysis.

If there is low energy supersymmetry, the determination of the
quantum numbers of the predicted particles will require comprehensive
studies of complicated cascade decays, taking advantage of spin
correlations.  If there is no low energy supersymmetry, physics beyond
the standard model will contribute to processes like $W$/$Z$ couplings
in~$e^+e^-\to f_1\bar f_2 f_3\bar f_4$, $WW$ scattering in~$e^+e^-\to
\nu_e\bar\nu_e f_1\bar f_2 f_3\bar f_4$, $t\bar t$ production
in~$e^+e^-\to f_1\bar f_2 f_3\bar f_4 f_5\bar f_6$ and $WW$/$t\bar t$
scattering in~$e^+e^-\to \nu_e\bar\nu_e f_1\bar f_2 f_3\bar f_4
f_5\bar f_6$, but the effects will be very small and can only be
observed in \emph{precision measurements} of these processes---and
complementary processes at hadron colliders---that must be compared to
\emph{precision calculations} in the standard model, effective field
theories and specific models for physics beyond the standard model.

Therefore, we will need reliable predictions, implemented in precision
simulation tools to unleash the full potential of future colliders.
Since the primary objective of the next generation of colliders is the
determination of the nature of electroweak symmetry breaking,
particular care must be taken to obtain gauge invariant predictions.
Polarization will be important to focus on longitudinal gauge bosons
and for measuring quantum numbers of supersymmetric particles.

It will be qualitatively more complicated to reach reliable high
precision predictions for these multi particle final states than for
the two particle final states that have dominated physics at LEP1: 1)~the
parameter space of models beyond the standard model is much larger,
2)~the number of Feynman diagrams contributing to a process explodes
combinatorially, 3)~algebraic expressions become much more
complicated with the growing number of building blocks
(i.\,e.~independent momenta and polarization vectors), 4)~gauge
cancellations become numerically hazardous, and 5)~phase space
also becomes much more intricate.

Therefore, even if we had enough graduate students and postdocs, we
should not waste them on repetitive ``assembly line'' calculations.
Instead, efforts must be directed toward formalizing the calculations
so that the repetitive part can be delegated to computers, which are
more patient with and particularly reliable for repetitive work. Of
course, this is not a new observation and has formed the rationale for
the AIHEP workshops starting in 1990, which were the precursor of the
present ACAT conferences.

In a bird's eye's view, the status and trends at ACAT\,2002 were:
1) tree-level calculations in the standard model for $2\to4$ and
$e^+e^-\to 6$ have been well under control for some time now and can
readily be used by non-experts, 2) the support for general $2\to6$ and
$2\to8$ processes is coming along, but not completely usable for
production yet (systems with already complete physics support do not
scale optimally, while optimally scaling systems have not been
implemented completely yet).

The main challenges lie beyond the standard model (in the MSSM in
particular) and virtual radiative corrections.  Regarding the former,
progress has been reported at the conference and regarding the latter,
there is growing evidence, concern and consensus that the classic
analytical approach ``does not scale''.  Recently, several projects
have started to work on different \mbox{(semi-)}numerical approaches and
progress by one group has been reported at this
conference~\cite{Pasarino}.

\section{Automated Calculation and Simulation Systems}
\begin{figure}
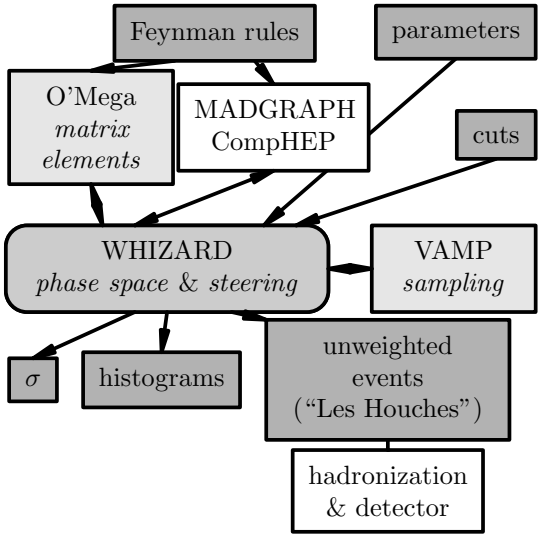

  \begin{center}
    \begin{empcmds}
      def vconnect (suffix from, ff, to, tf) =
  	(ff[from.sw,from.se]{down}---{down}tf[to.nw,to.ne]
  	 cutbefore bpath.from cutafter bpath.to)
      enddef;
      def hconnect (suffix from, ff, to, tf) =
  	(ff[from.ne,from.se]{right}---{right}tf[to.nw,to.sw]
  	 cutbefore bpath.from cutafter bpath.to)
      enddef;
      def fixsizepos (text t) =
  	fixsize (t);
  	fixpos (t)
      enddef;
    \end{empcmds}
    \begin{emp}(70,70)
      ahlength := 3mm;
      ahangle := 20;
      pickup pencircle scaled 1.5pt;
      defaultdx := 6pt;
      defaultdy := 6pt;
      boxit.feynman (btex Feynman rules etex);
      boxit.params  (btex parameters etex);
      boxit.cuts  (btex cuts etex);
      boxit.omega (btex \parbox{5em}{\centering
                          O'Mega\\ \emph{matrix elements}} etex);
      boxit.madgraph (btex \parbox{6em}{\centering
                             MADGRAPH\\ CompHEP} etex);
      rboxit.whizard (btex \parbox{11em}{\centering
                              WHIZARD\\ \emph{phase space} \&\ \emph{steering}} etex);
      boxit.vamp (btex \parbox{5em}{\centering VAMP\\ \emph{sampling}} etex);
      boxit.sigma (btex $\sigma$ etex);
      boxit.hist (btex histograms etex);
      boxit.events (btex \parbox{8em}{\centering unweighted\\ events\\ (``Les Houches'')} etex);
      boxit.jetset (btex \parbox{6em}{\centering hadronization\\ \&\ detector} etex);
      feynman.nw = (.2w,h);
      params.ne = (w,h);
      cuts.ne = (w,.8h);
      omega.w = (0, whatever);
      omega.c = .6[whizard.c,feynman.c] + (whatever,0);
      omega.e = madgraph.w - (.01w,0);
      whizard.c = (.3w,.5h);
      vamp.e = (w,.5h);
      events.c = .5[whizard.c,jetset.c] + (whatever,0);
      events.c = jetset.c + (0,whatever);
      hist.c = .5[sigma.e,events.w];
      events.c = sigma.c + (whatever,0);
      sigma.w = (0,whatever);
      jetset.se = (.9w,0);
      fixsizepos (whizard, feynman, params, cuts, sigma, hist,
                  events, jetset, vamp, omega, madgraph);
      fill bpath whizard withcolor .8white;
      drawboxed (whizard);
      fill bpath feynman withcolor .7white;
      fill bpath params withcolor .7white;
      fill bpath cuts withcolor .7white;
      drawboxed (feynman, params, cuts);
      fill bpath sigma withcolor .7white;
      fill bpath hist withcolor .7white;
      fill bpath events withcolor .7white;
      drawboxed (sigma, hist, events);
      fill bpath jetset withcolor white;
      drawarrow (vconnect (events, .5, jetset, .5));
      drawboxed (jetset);
      fill bpath whizard withcolor .8white;
      drawboxed (whizard);
      drawarrow (vconnect (params, .5, whizard, .8));
      drawarrow (vconnect (cuts, .5, whizard, .9));
      drawarrow (vconnect (whizard, .4, sigma, .5));
      drawarrow (vconnect (whizard, .5, hist, .5));
      drawarrow (vconnect (whizard, .7, events, 0));
      fill bpath vamp withcolor .9white;
      drawboxed (vamp);
      drawdblarrow (hconnect (whizard, .5, vamp, .5));
      fill bpath omega withcolor .9white;
      drawboxed (omega);
      drawarrow (vconnect (feynman, .3, omega, .5));
      drawdblarrow (vconnect (omega, .5, whizard, .3));
      fill bpath madgraph withcolor white;
      drawboxed (madgraph);
      drawarrow (vconnect (feynman, .7, madgraph, .5));
      drawdblarrow (vconnect (madgraph, .5, whizard, .4));
      setbounds currentpicture to (0,.15h)--(w,.15h)--(w,h)--(0,h)--cycle;
    \end{emp}
  \end{center}
  \caption{\label{fig:whizard-schema}%
    Components of the \texttt{WHIZARD} ACSS.  \texttt{WHIZARD}
    can use different components for generating matrix elements and
    relies on the \texttt{VAMP} library for adaptive multi channel
    sampling of phase space parameterizations matching the
    matrix element's dominant peaking structures.  The other ACSS are
    structured in a similar way.}
\end{figure}
\emph{Fully} Automated Calculation and Simulation Systems (ACSS)
in particle physics aim to produce cross sections and event samples
for a given physics model and scattering process
\begin{multline*}
  \left\{\begin{array}{cc}
           \text{model}      & \mathcal{L} \\
           \text{parameters} & m, g, \ldots \\
           \text{process}    & e^+e^-\to\nu_e\bar\nu_e\mu^+\nu_\mu d\bar u \\
           \text{cuts}       & p_{T,\text{min}}, E_{\text{min}}, \ldots
         \end{array} \right\} \\ \Longrightarrow
  \left\{\begin{array}{c}
           \sigma \\
           \text{event samples}
         \end{array} \right\}
\end{multline*}
with as little as possible human intervention (ideally without any,
of course).  The job of ACSS can roughly be be divided into two steps
\begin{enumerate}
  \item calculate the matrix element~$T$ (i.\,e.~generate Feynman
    diagrams, derive arithmetical expression and generate executable
    code)
  \item integrate $|T|^2$ with a phase space measure $\mathrm{d}\mu$
    or generate events according to~$|T|^2\mathrm{d}\mu$ (the
    efficiency of this step can be improved significantly by using
    information on the structure of~$T$).
\end{enumerate}
In a third step, the partonic final states have to be transformed into
hadronic final states.  This fragmentation and hadronization is
typically delegated to the established players in this field like
\texttt{Pythia} and \texttt{HERWIG}.

Still, an ACSS is a complex piece of software and can only be
constructed and maintained if broken into several components.  As a
typical example, the structure of \texttt{WHIZARD} is displayed in
figure~\ref{fig:whizard-schema}. Indeed, some systems are complete and
self-contained, while
some provide components of complete systems\footnote{Lack of space
makes it impossible to provide complete references. The ordering proceeds
from complete systems to components and is alphabetical in each group.}:
\texttt{CompHEP} is a complete system (two talks and two posters at ACAT\,2002),
\texttt{CalcHEP} is a \texttt{CompHEP} clone (one talk),
\texttt{GRACE} is a complete system (one talk and additional talks in the
sessions on numerical integration and sampling),
\texttt{O'Mega} calculates efficient matrix elements for many particle
  final states,
\texttt{WHIZARD} performs phase space integration for many particle
  final states and integrates other components into a complete system,
\texttt{HELAC/PHEGAS} calculates efficient standard model matrix
  elements and performs phase space integration,
\texttt{Madgraph} calculates standard model matrix elements\footnote{A
  companion program \texttt{MadEvent}~\cite{MadEvent} for event
  generation was introduced not too long after the conference.},
\texttt{Alpha} calculates efficient standard model matrix elements for many
  particle final states,
\texttt{FeynArts/\texttt{FeynCalc}/\texttt{FormCalc}} calculates
  standard model and MSSM loop diagrams,
\texttt{CalcPHEP} (now \texttt{SANC}) calculates standard model $2\to2$
  processes at one-loop (two talks and two posters).

\section{Trends at ACAT\,2002}
\subsection{Supporting More General Models}
Currently, all ACSS use a set of Feynman rules (i.\,e.~a set of
propagators and vertices) as input.  This is mathematically equivalent
to the Lagrangian describing the model.  However, the derivation of
the Feynman rules derivation from a Lagrangian---even if
straightforward---can be extremely tedious and error-prone
(e.\,g.~there are many thousand different vertices in the MSSM).

In two talks~\cite{Kaneko,Semenov}, tools for the derivation of Feynman
rules in complicated models---the MSSM in particular---have been
discussed at ACAT\,2002.  Such tools need to be able to 
expand lagrangians in terms of component fields and momenta and 
simultaneously handle mixing from non-diagonal mass matrices.
These operations must use a very natural notation in order to minimize
the potential for human errors.

Therefore, such tools~\cite{Kaneko,Semenov} evolve quickly into fully
developed symbolic manipulation languages, with some special features:
1)~transparent notation for objects from mutually commutative, but
non-commutative algebras, like non-simple gauge groups (this is
poorly supported in general purpose computer algebra systems) and 2)~support
for the dual role played by covariant derivatives as operators and
field components.  Soon there will be several \emph{independently}
derived sets of machine readable Feynman rules for the MSSM.

In addition, moving beyond the standard model requires that some
systems have to relax assumptions that were useful for implementing
the Standard Model in the first generation of ACSS:
Majorana spinors have to be supported in addition to Dirac spinors,
vertices of degree higher than four, more complicated momentum
dependence in higher dimensional operators, non-local vertices factors
(e.\,g.~in non-commutative field theories), non Feynman diagram
contributions (e.\,g.~K-matrix unitarization), etc.

One solution is to replace previously hard-coded subsystems making
special assumptions by general purpose components. One
poster reported how \texttt{CompHEP}'s special purpose
\texttt{C}-routines for the symbolic evaluation of squared diagram are being
replaced by calls to the general purpose \texttt{form} system.

\subsection{Component Architecture and Persistency}
The components in an ACSS have
to communicate across a variety of boundaries:
\textit{abstraction layers} (subroutines, objects, modules, etc.),
\textit{address spaces} (processes, networks, etc.),
and \textit{time} (data storage, persistency).
These issues were addressed by two talks~\cite{Lonnblad,Cherstnev}.

\texttt{Pythia7}~\cite{Lonnblad} is not an ACSS in the
the strict sense, but all ACSS have to talk to
\texttt{PYTHIA} or \texttt{HERWIG} for fragmentation and hadronization
of hard events. \texttt{Pythia7} is a project to completely rewrite the
Lund event generators in \texttt{C++}.  It will provide a general
structure for implementing models for event generation, \emph{not}
only the Lund Model (in fact, \texttt{HERWIG++} has joined the
effort). Today, \texttt{Pythia7} exists as a proof-of-concept version
(some basic $2\to2$ matrix elements, remnant handling and Lund string
fragmentation.  A new pre-release with the first \texttt{HERWIG}
contributions is planned for 2002 and a usable generator is planned
for the following year. \texttt{Pythia7} should be available as the
standard generator at the LHC start-up.

External representations of event samples are required for
communicating partonic events to soft QCD Monte Carlos and to detector
simulations.  External representations are also useful for saving both
intermediate and final results.  The latter can be useful if reading
the stored events is faster than re-generation.  However, there will
always be a gain for reweighting processed data and one must also not
underestimate the value of external representations for debugging.

A proposal of a text based data description language for event samples
and the implementation of the accompanying toolkit~\cite{Cherstnev}
was welcomed, but during the discussion there was demand for
replacing the concrete syntax by XML, while retaining the toolkit
interface, was voiced.

\subsection{Loops}
In two talks~\cite{Bardin,Christova} and two
posters, the project of a re-implementation
of the complete standard model one-loop radiative corrections
(incl.~soft bremsstrahlung) for~$e^+e^-\to f\bar f$ from scratch was
presented.

One motivation for such a re-implementation lies in the need to
preserve the body of knowledge for future generations by providing a
consistent and systematic option for redoing the calculations.  An
application will be the relaxation of $Z$-pole assumptions for
applications to off-resonance physics at a linear collider.

The calculational procedure starts by a decomposition into form
factors, followed by the calculation of the form factors using a
collection of \texttt{form3} procedures for symbolic calculation,
renormalization and generating numerical Fortran code.  Currently, the
$t$-channel neutral current processes and and the $s$-channel
processes are done and show very good agreement with existing codes.
Interference contributions and decays are in progress.

\begin{figure}
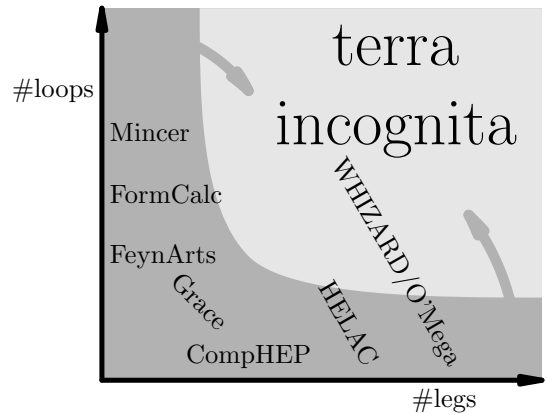

  \begin{center}
    \begin{emp}(65,55)
      fill (.1w,.1h)--(w,.1h)--(w,h)--(.1w,h)--cycle withcolor .7white;
      fill (.3w,h){down}...(.4w,.4h)...{right}(w,.3h)--(w,h)--cycle withcolor 0.9white;
      pickup pencircle scaled 4pt;
      drawarrow (.15w,.95h){right}...(.4w,.8h) withcolor .7white;
      drawarrow (.95w,.15h){up}...(.85w,.5h) withcolor .7white;
      pickup pencircle scaled 2pt;
      drawarrow (.1w,.1h)--(.1w,h);
      label.lft (btex \#loops etex, (.1w,.8h));
      label.rt (btex FeynArts etex, (.1w,.4h));
      label.rt (btex FormCalc etex, (.1w,.55h));
      label.rt (btex Mincer etex, (.1w,.7h));
      drawarrow (.1w,.1h)--(w,.1h);
      label.bot (btex \#legs etex, (.8w,.1h));
      label.top (btex Grace etex rotated -45, (.3w,.2h));
      label.top (btex \strut CompHEP etex, (.4w,.1h));
      label.top (btex \strut HELAC etex rotated -60, (.6w,.1h));
      label.top (btex \strut WHIZARD/O'Mega etex rotated -60, (.7w,.1h));
      label (btex \Huge\parbox{7em}{\centering terra\\incognita} etex, (.7w,.8h));
      setbounds currentpicture to (0,.15h)--(w,.15h)--(w,h)--(.1w,h)--cycle;
    \end{emp}
  \end{center}
  \caption{\label{fig:terra-incognita}%
    Schematic representation of the status of fully automated
    calculation tools.  (The names are a random representative of a
    class of programs.)  Important contributions can be
    made in the ``terra incognita'', that is covered neither by
    the existing many-loops/few-legs nor the many-legs/no-loops programs.}
\end{figure}

\section{Outlook}
If we would ever manage to produce a fully general and
efficient ACSS, that is even easy to use for non-experts,
we would risk to make ourselves obsolete as
phenomenological theorists.
Fortunately, we will never get there 
because theoretical theorists will always come up with
new features that we have not anticipated and experimentalists will
always push the frontiers in energy and precision, calling for ever
more precise calculations of ever more complicated processes.

Currently there is a vast uncharted territory
(figure~\ref{fig:terra-incognita}) and a systematic exploration of
the high energy/high precision region with many legs and many
loops will require the development of new semi-numerical methods.

Finally, variety is good, since we need to be able to cross check
our results.  Communication among developers and among programs---using
pluggable components in the ideal case---is necessary for this.

\subsection*{Acknowledgment}
I thank the organizers of ACAT\,2002 for the warm hospitality extended
to us at Moscow.


\end{empfile}
\end{document}